\documentclass[prb,twocolumn,showpacs,amsmath,amssymb,superscriptaddress]{revtex4-2}
\usepackage{dcolumn}
\usepackage{bm,graphicx}
\usepackage{etoolbox}
\usepackage{url}
\usepackage{booktabs}
\usepackage[colorlinks]{hyperref}
\usepackage{breakurl}
\usepackage{color}
\usepackage [latin1]{inputenc}

\usepackage{amsmath,amssymb}
\usepackage{xcolor}

\begin{document}

\title{High-angular momentum excitations in collinear antiferromagnet FePS$_3$}

\author{Jan Wyzula}
\affiliation{Laboratoire National des Champs Magn\'etiques Intenses, EMFL, CNRS UPR3228, Univ. Grenoble Alpes, Univ. Toulouse, Univ. Toulouse 3, INSA-T, Grenoble and Toulouse, F-38042, France}
\affiliation{Department of Physics, University of Fribourg, Chemin du Mus\'ee 3, CH-1700 Fribourg, Switzerland}
\author{Ivan Mohelsk\'y}
\affiliation{Laboratoire National des Champs Magn\'etiques Intenses, EMFL, CNRS UPR3228, Univ. Grenoble Alpes, Univ. Toulouse, Univ. Toulouse 3, INSA-T, Grenoble and Toulouse, F-38042, France}
\author{Diana V\'aclavkov\'a}
\affiliation{Laboratoire National des Champs Magn\'etiques Intenses, EMFL, CNRS UPR3228, Univ. Grenoble Alpes, Univ. Toulouse, Univ. Toulouse 3, INSA-T, Grenoble and Toulouse, F-38042, France}
\author{Piotr Kapuscinski}
\affiliation{Laboratoire National des Champs Magn\'etiques Intenses, EMFL, CNRS UPR3228, Univ. Grenoble Alpes, Univ. Toulouse, Univ. Toulouse 3, INSA-T, Grenoble and Toulouse, F-38042, France}
\author{Martin Veis}
\affiliation{Institute of Physics, Charles University, Ke Karlovu 5, Prague, CZ-121 16, Czech Republic}
\author{Cl\'ement~Faugeras}
\affiliation{Laboratoire National des Champs Magn\'etiques Intenses, EMFL, CNRS UPR3228, Univ. Grenoble Alpes, Univ. Toulouse, Univ. Toulouse 3, INSA-T, Grenoble and Toulouse, F-38042, France}
\author{Marek Potemski}
\affiliation{Laboratoire National des Champs Magn\'etiques Intenses, EMFL, CNRS UPR3228, Univ. Grenoble Alpes, Univ. Toulouse, Univ. Toulouse 3, INSA-T, Grenoble and Toulouse, F-38042, France}
\affiliation{CENTERA Labs, Institute of High Pressure Physics, PAS, PL-01-142 Warsaw, Poland}
\author{Mike E. Zhitomirsky}
\affiliation{Univ. Grenoble Alpes, CEA, IRIG, PHELIQS, 17 avenue des Martyrs, F-38000 Grenoble, France}
\author{Milan Orlita}
\affiliation{Laboratoire National des Champs Magn\'etiques Intenses, EMFL, CNRS UPR3228, Univ. Grenoble Alpes, Univ. Toulouse, Univ. Toulouse 3, INSA-T, Grenoble and Toulouse, F-38042, France}
\affiliation{Institute of Physics, Charles University, Ke Karlovu 5, Prague, CZ-121 16, Czech Republic}
\email{milan.orlita@lncmi.cnrs.fr}

\begin{abstract}
  We report on magneto-optical studies of the quasi-two-dimensional
van der Waals
antiferromagnet FePS$_3$. Our measurements reveal an excitation that closely resembles the antiferromagnetic resonance mode typical of easy-axis antiferromagnets, nevertheless, it displays an unusual, four-times larger Zeeman splitting
in an applied magnetic field. We identify this excitation with
an $|S_z|=4$ multipolar
magnon -- a single-ion 4-magnon bound state -- that corresponds to a full reversal of a single magnetic moment of the Fe$^{2+}$ ion.
We argue that condensation of multipolar magnons in large-spin materials with a strong magnetic anisotropy can produce new exotic states.
\end{abstract}

\maketitle

Magnons or quantized spin waves are collective excitations of localized magnetic moments in crystalline solids. These boson-like
quasiparticles disperse with a lattice momentum $\mathbf{k}$, carry a fixed amount of energy and have an integer spin or angular momentum $S_z=\pm 1$.
Magnons are optically active excitations, though due to vanishing momentum of photons, only spatially uniform spin waves are excited ($k=0$ magnons).
Despite the strictly quantum nature of magnons, their optical and magneto-optical response can be approached at the semi-classical
level~\cite{KittelPR48,KittelPR51}, as transverse precession of coupled magnetic dipoles. In antiferromagnets, the related resonant absorption of light is
referred to as the antiferromagnetic resonance (AFMR).

\begin{figure}[t]
     \includegraphics[width=0.37\textwidth]{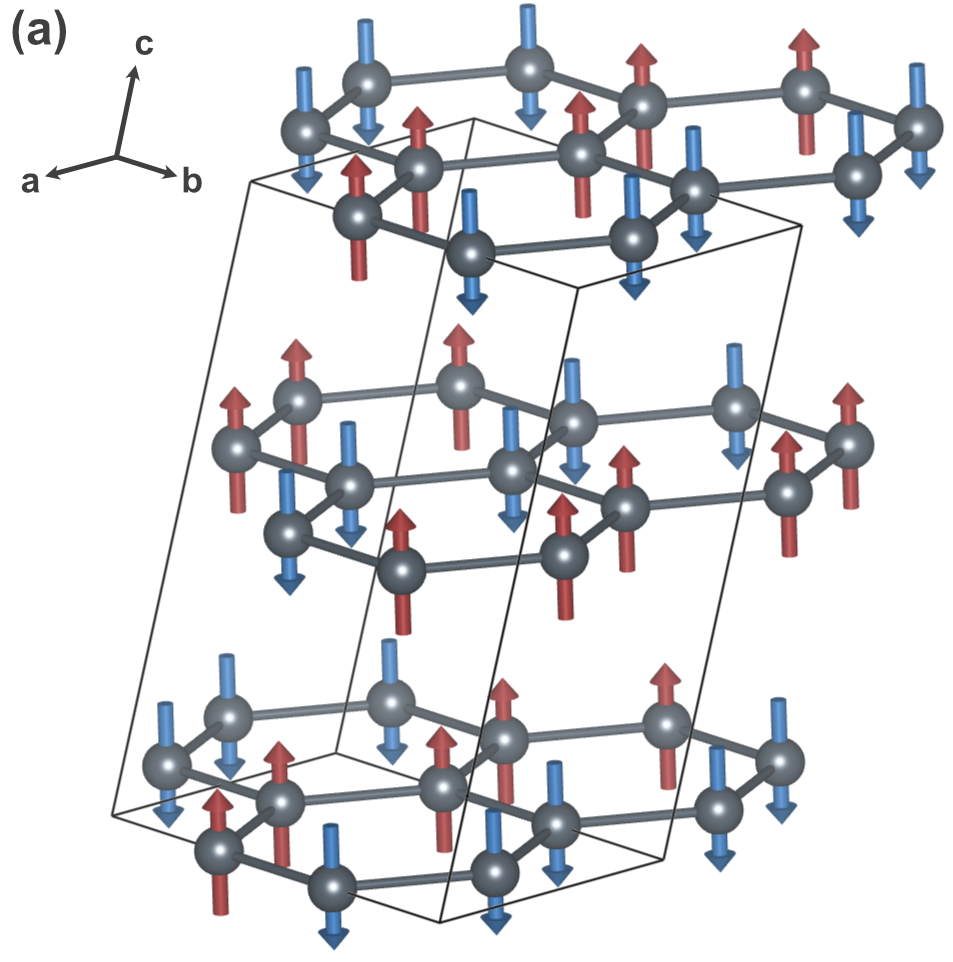}
      \caption{\label{Lattice}  The crystal lattice of FePS$_3$. The magnetic moments 
      are color-coded to differentiate between the two orientations. The figure was created using the VESTA software~\cite{VESTA}.}
\end{figure}

The AFMR has a particularly simple form in an antiferromagnet with a uniaxial anisotropy. Once an external magnetic field $B$ is applied along the easy axis, the single AFMR mode splits into two frequency branches that are linear in $B$~\cite{KittelPR51,KefferPR52}:
\begin{equation}
\omega_{\mathrm{AF}} = \omega_{\mathrm{AF}}^{0} \pm \gamma B,
\label{Kittel}
\end{equation}
where $\gamma$ stands for the gyromagnetic ratio.
This implies the separation of branches equal to twice the Larmor frequency, or equivalently, to twice the Zeeman energy, 
$2E_Z = 2 g \mu_B B$, with the $g$ factor typically close to the free-electron value ($g=2$).

The AFMR that follows such a simple rule was observed in the GHz/THz frequency range
for many easy-axis antiferromagnets: in iron and manganese dihalides~\cite{PetitgrandJdP76,HagiwaraIJIMW99} or in magnetic oxides~\cite{EllistonJPC68,FowlisCJP72,MihalyPRB04,ZHANGJMMM21}, to name a few. At present, the AFMR in easy-axis antiferromagnets represents perhaps the most characteristic and easy-to-identify resonance/excitation in magnetic systems and it serves as a direct probe of the exchange coupling between spins and of the single-ion magnetic anisotropy.

\begin{figure*}[t!]
     \includegraphics[width=0.95\textwidth]{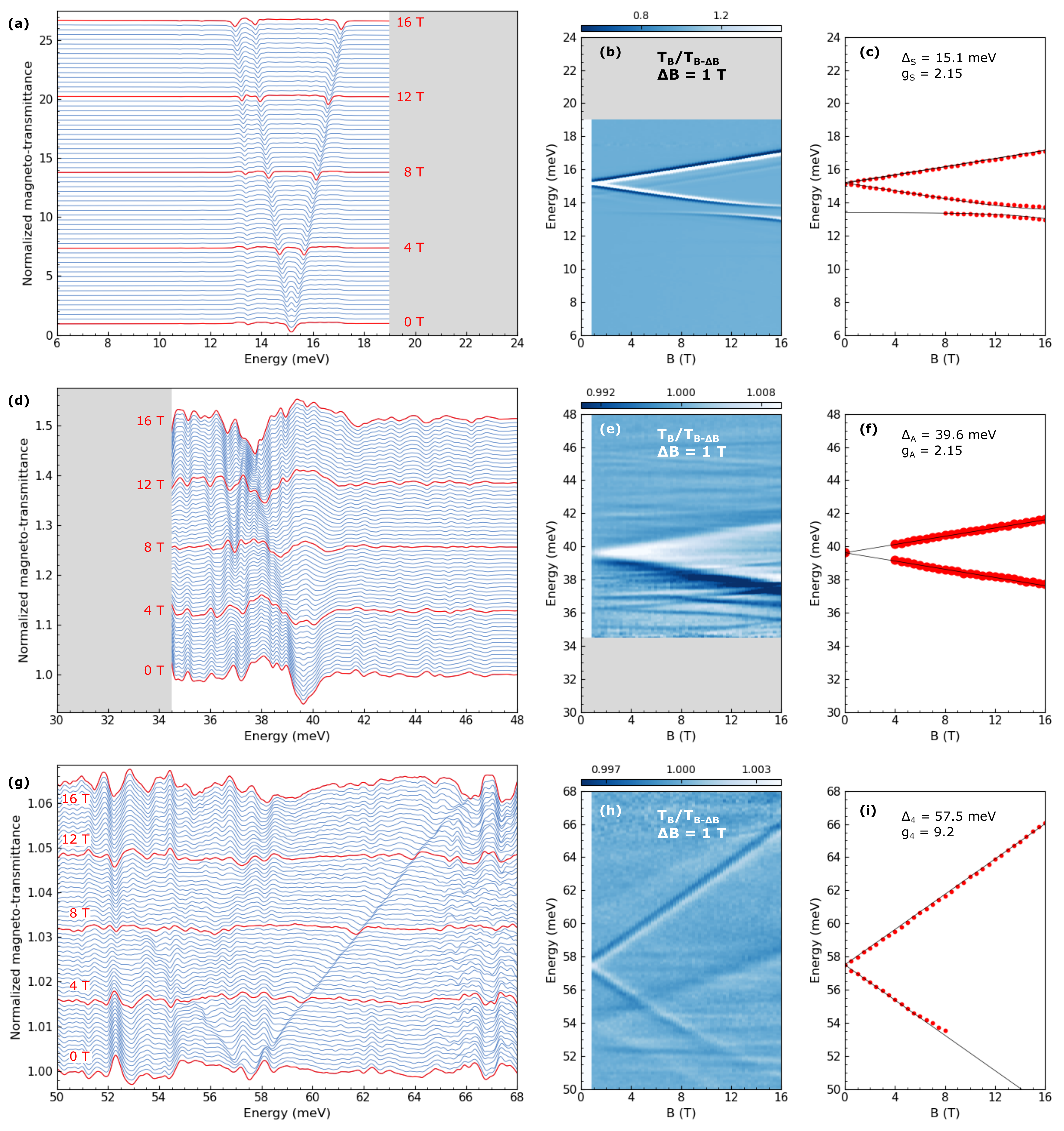}
      \caption{\label{Fig} Magneto-transmission of FePS$_3$ at $T=4.2$~K in
      three spectral ranges showing the AFMR due to the lower 1-magnon gap $\Delta_S$ (a-c), the upper 1-magnon gap $\Delta_S$ (d-f) and the 4-magnon mode $\Delta_4$ (g-i). The magneto-transmission spectra in (a), (d) and (g) were normalized by the averaged transmission $T_B$ average over the explored interval $B=0-16$~T. In panels (b), (e) and (h), differential magneto-transmission is plotted, $T_B/T_{B-\Delta B}$ for $\Delta B = 1$~T and the step in $B$ of 0.25~T. The extracted position of all magnon branches (red circles) at selected $B$ is presented in panels (c), (f) and (i) along with the theoretical fit (black lines). In panel (c), the theoretical fit uses the model and parameters described in Ref.~\onlinecite{VaclavkovaPRB21}. In panels (f) and (i), the fit assumes two linear-in-$B$ AFMR-like branches. In gray areas, the explored FePS$_3$ sample was opaque.}
\end{figure*}

In this Letter, we report on THz/infrared magneto-spectroscopy studies of the quasi-two-dimensional honeycomb antiferromagnet FePS$_3$, see Figure~\ref{Lattice},
which belongs to a topical family of magnetic van der Waals materials~\cite{Park16}.
We find that the response is dominated by three types of magnetic excitations. Two of them correspond to conventional 1-magnon gaps which may, when the magnetic field is applied, be taken as text-book examples of the AFMR in easy-axis antiferromagnets. At the qualitative level, the third mode also resembles the AFMR, nevertheless, it displays four-times larger splitting of the branches. We show that this excitation corresponds to a full reversal of a single spin $S=2$, which carries a total angular momentum of $|S_z|=4$. Our observation extends the concept of single-ion bound states, so far limited to 2-magnon single-ion bound states in $S=1$ systems~\cite{SilberglittPRB70}, towards more complex excitations with a multipolar symmetry. In particular, for $S=2$ magnetic materials, their condensation in a magnetic field may allow for exotic hexadecapole phases. Overall, our findings illustrate an emergence of exotic quantum excitations in semiclassical magnetic materials with large spins.

The infrared magneto-transmission experiments were carried out on a macroscopic FePS$_3$ bulk sample (with a surface area of~10~mm$^2$) which was kept in helium exchange gas at temperature of 4.2~K and placed in a superconducting solenoid. The magnetic field was always applied normal to the quasi-two-dimensional planes of FePS$_3$ and aligned with the wave vector of the probing radiation (Faraday configuration). To measure magneto-transmission, the radiation from globar was analyzed by a Bruker Vertex 80v Fourier-transform spectrometer, delivered to the sample via light-pipe optics and detected by a composite bolometer placed directly behind the sample. All presented data were collected using the spectral resolution of 0.125~meV.

The collected magneto-transmission data are presented in Figure~\ref{Fig} in three relevant spectral windows. To allow for a detailed inspection of these data, they are plotted in the form of magneto-transmission $T_B$, normalized by the transmission averaged over the whole range of $B$ explored, in panels (a), (d) and (g), and also as false-color plots of differential magneto-transmission, $T_B/T_{B-\Delta B}$,  in panels (b), (e) and (h). In each chosen window, the response is dominated by a single magnetic mode that splits into two linear-in-$B$ branches, thus resembling closely the AFMR. These modes, denoted as $M_S$, $M_A$ and $M_4$, are centered at the photon energies of $\Delta_S=$~15.1~meV, $\Delta_A=$~39.6~meV and $\Delta_4=$~57.5~meV, respectively. The deduced positions of both AFMR branches are plotted in parts (c), (f) and (i), respectively, as a function of
$B$. The $M_4$ mode has the integral intensity roughly by a factor of $10^3$ smaller as compared to $M_S$ and $M_A$ modes. Our data comprise even weaker spectral features, for instance, the AFRM-like modes at $\sim$54 and 61~meV visible in Figure~\ref{Fig}h, which are analyzed and discussed in the Supplementary materials~\cite{SM}.

The quasi two-dimensional antiferromagnet FePS$_3$ consists of honeycomb planes formed by Fe$^{2+}$ ions with $S=2$, see Figure~\ref{Lattice}. Below $T_N\approx 120$~K, FePS$_3$ orders in a collinear zig-zag structure with four spins per unit cell~\cite{RulePRB07}. Magnetic moments are oriented orthogonally to honeycomb layers indicating the easy-axis anisotropy. The excitation spectrum consists of two double-degenerate magnon branches that were observed in the inelastic  neutron scattering experiments~\cite{WildesJPCM12,LanconPRB16,WildesJAP20}. Comparison with the neutron
data allows us to  assign the $M_S$ and $M_A$ modes to 1-magnon ($S_z=\pm 1$) excitations 
with $k = 0$ corresponding to in-phase and out-of-phase oscillations of parallel magnetic sublattices, respectively.

The low-energy mode $M_S$ has been identified as the 1-magnon gap also in preceding  optical (Raman scattering) studies~\cite{SekinePRB90}. In a magnetic field, this mode splits into two AFMR branches (\ref{Kittel}) separated by twice the Zeeman energy~\cite{McCrearyPRB20}. The corresponding $g$ factor is $g_S \approx 2.15$ (Figure~\ref{Fig}c). At higher magnetic fields, the lower branch of the $M_S$ mode departs from a linear field dependence (Figures.~\ref{Fig}b,c). This is due to a significant magnon-phonon coupling
recently reported for FePS$_3$~\cite{LiuPRL21, VaclavkovaPRB21,ZhangCM21,PawbakeACSNano22}. The solid
line in Figure~\ref{Fig}c that describes this coupling is a fit, using the model and its parameters discussed in Ref.~\onlinecite{VaclavkovaPRB21} in detail. The upper 1-magnon gap $M_A$ exhibits the same AFMR-like splitting, $g_A=g_S$,
cf.~Figures~\ref{Fig}c and \ref{Fig}f. Interestingly, no traces of this upper 1-magnon gap have so far been found in Raman scattering measurements to the best of our knowledge, nevertheless, its existence is clearly evidenced, \emph{e.g.}, in neutron scattering experiments~\cite{LanconPRB16}.

To develop a quantitative description  of the observed resonance modes, we use the spin model for FePS$_3$ based on the neutron experiments~\cite{WildesJPCM12,LanconPRB16}. The spin Hamiltonian of FePS$_3$ includes the Heisenberg exchange interactions up to the third neighbours as well as the single-ion anisotropy and Zeeman terms:
\begin{equation}
\begin{split}
\hat{\mathcal H} = J_1 \sum_{\langle ij\rangle} {\bf S}_i\cdot {\bf S}_j +
J_2 \sum_{\langle ij\rangle} {\bf S}_i\cdot {\bf S}_j + J_3 \sum_{\langle ij\rangle} {\bf S}_i\cdot {\bf S}_j \\ - D \sum_i (S_i^z)^2 + g\mu_B B\sum_{i} S_i^z\,.
\end{split}
\label{Heisenberg}
\end{equation}
The zigzag antiferromagnetic order observed in FePS$_3$ is stable for a sufficiently strong antiferromagnetic third-neighbour coupling $J_3>0$~\cite{FouetEPJB01}. Let us emphasize that this Hamiltonian is strictly two-dimensional, thus neglecting any inter-layer exchange coupling between spins. Such an approach is corroborated by results of neutron scattering measurements~\cite{LanconPRB16}, but also Raman scattering experiments~\cite{LeeNL16,Wang2D16} which indicate that FePS$_3$ keeps antiferromagnetic ordering down to the monolayer thickeness, with the N\'eel temperature barely changed. Hence, bulk FePS$_3$ is a unique system that allows us to address, by means of optical tools, two-dimensional magnons -- excitations seemingly untraceable in a monolayer sample (see Refs.~\onlinecite{LeeNL16,Wang2D16}). 

The magneto-optical experiments probe uniform magnon modes with $k=0$. For large-$S$ antiferromagnets, the conventional single-magnon excitations can be obtained rather accurately by the harmonic spin-wave theory. Details of the corresponding calculations are presented in the Supplementary material~\cite{SM}.
The four uniform modes split into two pairs corresponding to the in-phase and out-of-phase oscillations of parallel sublattices in the up-up-down-down magnetic structure of FePS$_3$. The two symmetric modes have energies:
\begin{equation}
\Delta_S = 2S \sqrt{D(D+J_1+4J_2+3J_3)}\pm g \mu_B B \ ,
\label{MagnonS}
\end{equation}
whereas the antisymmetric modes are:
\begin{equation}
\Delta_A = 2S \sqrt{(D-2J_1+4J_2)(D-J_1+3J_3)} \pm g \mu_B B \ .
\label{MagnonA}
\end{equation}

In collinear antiferromagnetic structures with a rotation symmetry around the easy axis,
magnons have a well-defined quantum numbers $S_z=\pm 1$. An applied magnetic field splits every degenerate pair of modes into
descending and ascending branches with the bare $g$-factor, $g_{S,A}\approx 2.15$ in FePS$_3$. To get numerical estimates, we use the microscopic parameters obtained
by Lancon {\it et al.}~\cite{LanconPRB16} from experimental fits of
magnon bands (in meV):
\begin{equation}
J_1 = -2.92\,,\ \ J_2 = 0.08\,,\ \ J_3 = 1.92\,,\ \ D = 2.66\,.
\label{JD}
\end{equation}
Because of a different convention used for presenting the lattice sums in the spin Hamiltonian (\ref{Heisenberg}), the above values for the exchange constants $J_n$ differ by a factor of $-2$ from those given in Ref.~\onlinecite{LanconPRB16}. Using Eqs.~(\ref{MagnonS}) and (\ref{MagnonA}) we reproduce  the two lowest  resonance modes measured in our experiments with an accuracy of 1--3~\%.

Let us now consider the third mode $M_4$ which is observed at a high energy, $\Delta_4 = 57.5$~meV (at $B=0$). Notably, this energy is slightly smaller than four times the energy of the lowest magnon: $\Delta_4 \lesssim 4\Delta_S \approx 60.4$~meV. Under magnetic field, the $M_4$ mode splits into two components with the effective $g$-factor  $g_4\approx 4g_{S,A}$. Such a relation points at a 4-magnon nature of this extra mode. The strong easy-axis anisotropy observed in FePS$_3$ can provide a mechanism  for the creation of additional low-energy excitations besides the conventional magnon modes. Full reversal of a single spin, say, from $S_z = +S$ to $S_z=-S$ does not change the single-ion energy  and costs only the energy of broken exchange bonds. It carries a high angular momentum  $|S_z|=2S$, which in the case of FePS$_3$ amounts to $|S_z|=4$. The ordinary magnons have instead $|S_z|=1$ irrespective of the spin quantum number of constituent magnetic ions.
We shall use the term multipolar magnons for the full spin reversal excitations.  The conventional magnons correspond in this terminology to dipolar excitations.
A multipolar magnon can be viewed as $2S$ ordinary magnons
attracted to the same lattice site by the single-ion anisotropy.  Thus, this excitation can be alternatively named a single-ion bound state~\cite{SilberglittPRB70} in contrast to the exchange bound states discussed, for example, in Refs.~\onlinecite{Wortis63,Hanus63,OguchiJPSJ73,Chubukov91,Shannon06,HamerPRB09,MZH10}.

The interest in single-ion bound states, sometimes also referred to as longitudinal magnons,
has a long history. The focus of the majority of experimental~\cite{FertSSC78,BahurmuzPRB80,PetitgrandJMMM80,PsaltakisJPC84,OrendacPRB99,ZvyaginPRL07,ZvyaginPBCM08,BaiNP21} and theoretical~\cite{SilberglittPRB70,Papanicolaou87,DamlePRL06,WierschemPRB12,SizanovJETP13} studies was so far on $S=1$ magnets.
In spin-1 materials, 2-magnon bound states with $|S_z|=2$ may appear and lead to quadrupolar or spin-nematic instabilities~\cite{DamlePRL06,WierschemPRB12}.
More recently, the observations of bound states of more than two magnons have been reported.
The 3-magnon bound state was found in $\alpha$-NaMnO$_2$ ($S=2$) using the neutron scattering technique~\cite{DallyPRL20}, whereas the 4- and 6-magnon bound states hybridized with 1-magnon excitations were observed using the time-domain THz magneto-spectroscopy in FeI$_2$ ($S=1$)~\cite{LegrosPRL22}. Nevertheless, in both cases, the observed bound states are composed of bound magnons located on adjacent lattice sites, and therefore, correspond to the exchange bound states. Note that the multi-magnon bound states are to be distinguished from other excitations involving multiple 1-magnon states reported experimentally, see, \emph{e.g.}, Refs.~\onlinecite{DietzPRL71,CottamJPCSSP72,FunkenbuschSSC81}.

\begin{figure}[t]
\includegraphics[width=0.5\textwidth]{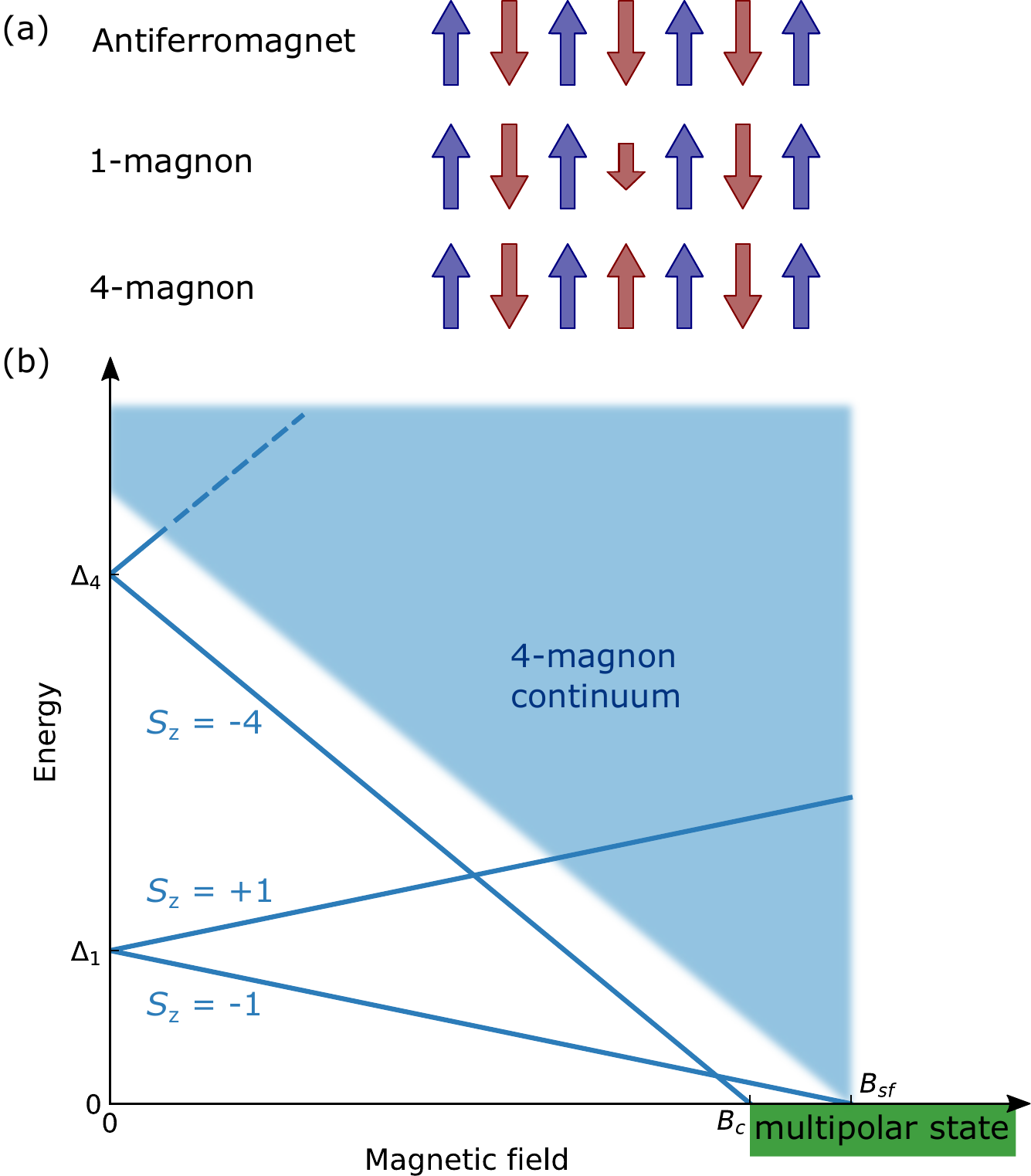}
\caption{Exotic excitations and states in a collinear $S=2$ antiferromagnet.
Panel (a): Schematic representation of a 1-magnon excitation
and a 4-magnon excitation. Panel (b): Resonance $k=0$ modes
in a magnetic field applied along the easy axis. $\Delta_1$ and $\Delta_4$
denote the AFMR and the 4-magnon gaps, respectively.
The condensation of the AFMR mode results in the usual spin flop transition
at $B=B_{\rm sf}$, whereas condensation of a 4-magnon at $B=B_c$
signifies appearance of a multipolar state.}
\label{Fig2}
\end{figure}

To corroborate our interpretation, let us estimate the energy of the $M_4$ mode.
In the crudest approximation, we neglect dispersion of a multipolar magnon
and compute its energy by inverting fully one spin in the zigzag antiferromagnetic 
structure of FePS$_3$~\cite{SM}:
\begin{equation}
\Delta_4 = 2S^2 (3J_3-J_1+2J_2)\pm 4g \mu_B B \,.
\label{EL}
\end{equation}
The two  branches, degenerate at $B=0$, correspond to spin flips on up and down sublattices.
The relation $g_4 = 4g$ remains exact for the continuous spin-rotation symmetry about the $z$ axis present in the spin Hamiltonian (\ref{Heisenberg}). A small deviation observed experimentally $g_4/g\approx 4.3$ can be attributed to weak spin-orbit effects. Notably, the above expressed 4-magnon energy -- corresponding to a full spin inversion -- does not depend on the $D$ parameter. This is because the magnetic anisotropy term in Hamiltonian (\ref{Heisenberg}) is even in $S_z$. In contrast, the energies of 1-magnon excitations (\ref{MagnonS},\ref{MagnonA}), and consequently, also various bound states of 1-magnons located on different lattice sites, always scale with $D$.

Using Eq.~(\ref{EL}) and the microscopic parameters (\ref{JD}), we obtain $\Delta^{\mathrm{th}}_4=71$~meV, which slightly deviates from the experimental value $\Delta_4=57.5$~meV. One of the effects not included in the theoretical expression (\ref{EL}) is the dispersion 
of the multipolar magnon. This problem can be approached in the strong-coupling limit $D\gg |J_n|$. For $S=2$, the transverse
coupling of full spin flips on adjacent sites appears only in the fourth-order of the perturbative expansion ${\cal J}_\perp  \simeq J_n^4/(8D^3)$
in comparison to the second-order effect for $S=1$~\cite{DamlePRL06,WierschemPRB12}. Since  in FePS$_3$
$D\sim |J_1|$, a more elaborated theoretical treatment is necessary to calculate the dispersion of a single-ion 4-magnon bound state. 

Nevertheless, the principal source of the discrepancy is likely related to the presence of a significant biquadratic exchange in FePS$_3$ as recently argued by Wildes {\it et al.}~\cite{WildesJAP20} from a detailed fit of the early inelastic neutron-scattering results. Obviously, such additional terms do not affect the multipolar magnon energy computed in the `Ising' approximation (\ref{EL}), whereas they
affect the 1-magnon dispersion, modifying an appropriate choice of the bilinear exchanges.
Taking instead of (\ref{JD}) the exchange constants $J_1=-2.92$~meV, $J_2=0.44$~meV and $J_3=1.1$~meV reported in Ref.~\onlinecite{WildesJAP20}, we obtain $\Delta^{\mathrm{th}}_4 =56.8$~meV in a perfect match with our experimental value.

Because of their high-angular momentum, the multipolar magnons should not be nominally
seen using experimental probes that obey the dipolar selection rules.
This corresponds to both the magneto-optical measurements and the neutron-scattering experiments. Our observation of such excitations is related to the fact that
FePS$_3$ has weak additional spin-orbit terms not included in Eq.~\ref{Heisenberg}
that break continuous rotation symmetry about the $z$ direction. When interlayer
coupling is neglected, we deal with a trigonal symmetry in FePS$_3$. As a result,
there is a weak hybridization between $S_z=4$ and $S_z=1$ excitations~\cite{SM},
which enables observation of multipolar magnons. Hence, even though the strength of the spin-orbit interaction -- quantified by the magnetic inisotropy $D$ -- does not directly enters the energy of the observed 4-magnon single-ion bound state (\ref{EL}), its role is still very important. First, it is responsible for a relatively high energy of the lower 1-magnon gap, and consequently, also for the high onset of the 
4-magnon continuum ($4\Delta_S$). This results in a relatively broad spectral window, in which the 4-magnon single-ion bound states can be searched. Second, it couples the local spin to the surrounding (orbital) crystal field thanks to which a multipolar magnon state can become optically active. Third, it stabilizes the 4-magnon single-ion  bound state.

We finish with a brief comment on possible new magnetic-field-induced states related to the presence of multipolar magnons.  For an easy-axis antiferromagnet with a moderate anisotropy
$D\sim J$ and an arbitrary spin $S\geq 1$, a multipolar magnon may split down from the $2S$-magnon continuum. The schematic frequency-field diagram is illustrated in Figure~\ref{Fig2} for a specific spin value $S=2$. The multipolar  magnon will condense~\cite{ZapfRMP14} in a magnetic field of $g_4\mu_B B_c =\Delta_4$ before a conventional spin-flop transition appears at $g\mu_B B_{\rm sf}=\Delta_1$. The nature of the high-field phase emerging at $B>B_c$ depends on the angular momentum of the condensed excitation. For $S=1$, this state breaks the quadrupolar symmetry and corresponds to a spin nematic phase~\cite{DamlePRL06,WierschemPRB12}. For larger spins $S>1$, higher-order multipoles are necessary for description of the symmetry-broken state. In particular, for $S=3/2$ and $S=2$, the corresponding order parameters have octupole and hexadecapole symmetry, respectively. Further details, for example, the instability wavevector at $B=B_c$, depend on the dispersion of multipolar magnons, which represents an interesting theoretical problem for future studies. Thus, contrary to conventional wisdom, which suggests a semiclassical behavior for magnets with $S>1$, the unconventional states arising from the condensation of multipolar magnons promise to be very interesting and exotic.

In conclusion, we have explored the magneto-optical response of the quasi-two-dimensional antiferromagnetic semiconductor FePS$_3$ and we have identified a novel magnetic excitation. It closely resembles the conventional AFMR, typical of easy-axis antiferromagnets, nevertheless, with a splitting four times larger than expected for 1-magnon gaps. We argue that we observe a multipolar excitation (single-ion 4-magnon bound state) which corresponds to a full reversal of a single Fe$^{2+}$ iron spin and carries a total angular momentum of $|S_z|=4$. This opens a venue for exotic hexadecapole phases.

\begin{acknowledgements}
We acknowledge useful discussions with Andrew Wildes. The work has been supported by the EU Graphene Flagship project. M.V.\ acknowledges the support by the Operational Program Research, Development and Education financed by European Structural and Investment Funds and the Czech Ministry of Education, Youth and Sports (Project MATFUN -- CZ.02.1.01/0.0/0.0/15\_003/0000487). M.E.Z.\ acknowledges support by the ANR,  France,  Grant No.~ANR-19-CE30-0004. The work was supported by the Czech Science Foundation, project No. 22-21974S and by the Czech-French exchange programme PHC Barrande (48101TB). This research was supported by the NCCR MARVEL, a National Centre of Competence in Research, funded by the Swiss National Science Foundation (grant number 205602).
\end{acknowledgements}


%

\newpage
\pagenumbering{gobble}

\begin{figure}[htp]
\includegraphics[page=1,trim = 17mm 17mm 17mm 17mm,width=1.0\textwidth,height=1.0\textheight]{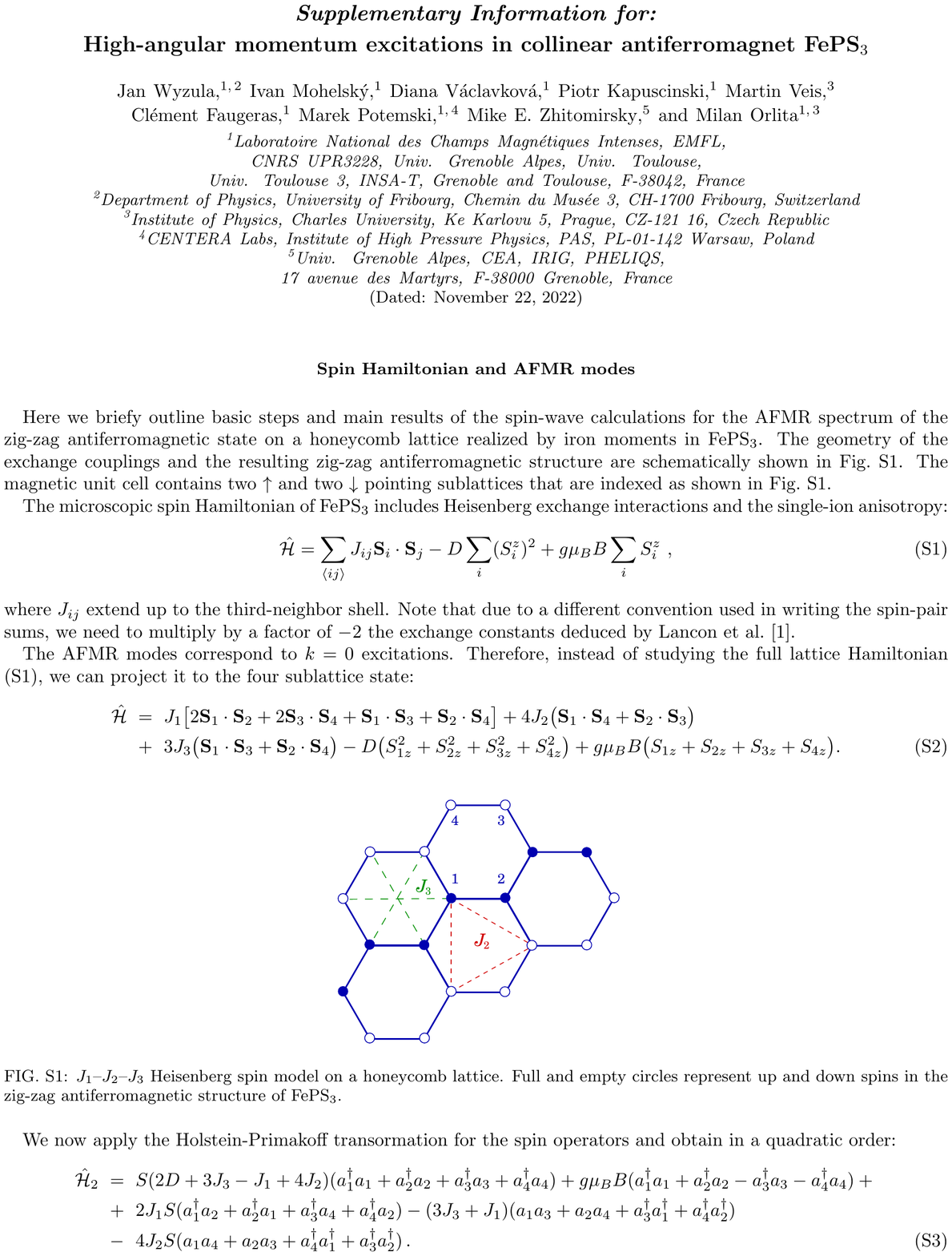}
\end{figure}

\newpage

\begin{figure}[htp]
\includegraphics[page=2,trim = 17mm 17mm 17mm 17mm, width=1.0\textwidth,height=1.0\textheight]{suppl.pdf}
\end{figure}

\newpage

\begin{figure}[htp]
\includegraphics[page=3,trim = 17mm 17mm 17mm 17mm, width=1.0\textwidth,height=1.0\textheight]{suppl.pdf}
\end{figure}

\end{document}